# Atomic clock performance beyond the geodetic limit

W. F. McGrew[1,2], X. Zhang[1,*], R. J. Fasano[1,2], S. A. Schäffer[1,†], K. Beloy[1], D. Nicolodi[1], R. C. Brown[1,‡], N. Hinkley[1,2,§], G. Milani[1,‖], M. Schioppo[1,¶], T. H. Yoon[1,**] and A. D. Ludlow[1]

[1]National Institute of Standards and Technology, 325 Broadway, Boulder, CO 80305, USA

[2]Department of Physics, University of Colorado, Boulder, CO 80309, USA

*Permanent address: State Key Laboratory of Advanced Optical Communication Systems and Networks, Institute of Quantum Electronics, School of Electronics Engineering and Computer Science, Peking University, Beijing 100871, China

†Permanent address: Niels Bohr Institute, University of Copenhagen, Blegdamsvej 17, 2100 Copenhagen, Denmark

‡Present address: Georgia Tech Research Institute, Atlanta, GA 30332, USA

§Present address: Stable Laser Systems, Boulder, CO 80301, USA

‖Permanent address: Istituto Nazionale di Ricerca Metrologica, Strada delle Cacce 91, 10135 Torino, Italy; Politecnico di Torino, Corso duca degli Abruzzi 24, 10125 Torino, Italy

¶Present address: National Physical Laboratory (NPL), Teddington, TW11 0LW, United Kingdom

**Permanent address: Department of Physics, Korea University, 145 Anam-ro, Seongbuk-gu Seoul 02841, South Korea

Correspondence should be addressed to A.D.L. (email: andrew.ludlow@nist.gov).

The passage of time is tracked by counting oscillations of a frequency reference, such as Earth's revolutions or swings of a pendulum. By referencing atomic transitions, frequency (and thus time) can be measured more precisely than any other physical quantity, with the current generation of optical atomic clocks reporting fractional performance below the $10^{-17}$ level[1–5]. However, the theory of relativity prescribes that the passage of time is not absolute, but impacted by an observer's reference frame. Consequently, clock measurements exhibit sensitivity to relative velocity, acceleration and gravity potential. Here we demonstrate optical clock measurements surpassing the present-day ability to account for the gravitational distortion of space-time across the surface of Earth. In two independent ytterbium optical lattice clocks, we demonstrate unprecedented levels in three fundamental benchmarks of clock performance. In units of the clock frequency, we report systematic uncertainty of $1.4 \times 10^{-18}$, measurement instability of $3.2 \times 10^{-19}$ and reproducibility characterised by ten blinded frequency comparisons, yielding a frequency difference of $[-7\pm(5)_{stat}\pm(8)_{sys}] \times 10^{-19}$. While differential sensitivity to gravity could degrade the performance of these optical clocks as terrestrial standards of time, this same sensitivity can be used as an exquisite probe of geopotential[5–9]. Near the surface of Earth, clock comparisons at the $1 \times 10^{-18}$ level provide 1 cm resolution along gravity, outperforming state-of-the-art geodetic techniques. These optical clocks can further be used to explore geophysical phenomena[10], detect gravitational waves[11], test general relativity[12] and search for dark matter[13–17].

Einstein first predicted in his general theory of relativity that gravity alters time, an effect sometimes called the gravitational redshift. Relative to a given observer, time (and the devices that measure time – clocks) are seen to evolve more slowly deeper in a gravity potential. To make meaningful comparisons between atomic clocks, this shift must be accounted for by transforming into a common reference frame, such as the geoid, which is the equipotential surface that best fits global mean sea level of the rotating Earth. In practice, the internationally recognized coordinate time system Terrestrial Time (TT) uses a reference frame corresponding to a surface of constant geopotential that is near the geoid but does not change due to, for example, eustatic sea level rise[18]. Relative heights between two nearby (as much as a few hundred km) locations can be determined with millimetre resolution by spirit levelling[19], but absolute geopotential determination can best be performed by using the Global Navigation Satellite System (GNSS) to measure ellipsoidal height (<1 cm uncertainty) and accounting for gravity by using a geoid model (several cm uncertainty)[20,21]. Near the surface of Earth, relativistic effects amount to a fractional frequency shift of $1.1 \times 10^{-18}$ per cm of vertical displacement. The geopotential determination afforded by modern GNSS and gravity measurements is sufficient for state-of-the-art microwave clocks with a systematic uncertainty approaching $1 \times 10^{-16}$, corresponding to 0.9 m of elevation change, but the next generation of clocks has the potential to push the boundaries of geodetic precision. These clocks are based on optical transitions, in which greater oscillation frequency leads to an increase of $10^5$ in the quality factor of the transition and concomitant performance improvements[22]. Clock performance is characterised by systematic uncertainty (the potential deviation of the measured transition frequency from its unperturbed value), instability (the statistical precision that the clock affords a measurement) and reproducibility (the measured agreement between similar but distinct clocks). We report the realization of unprecedented levels in each of these three benchmarks and demonstrate that these clocks can now perform beyond the present-day ability to account for gravitational effects of time across the surface of Earth. This opens the possibility for using clocks as precise, next-generation geodetic tools.

**Systematic uncertainty.** We characterise all known sources of systematic uncertainty in two ytterbium optical lattice clocks, denoted Yb-1 and Yb-2 throughout (Fig. 1). We report a total uncertainty of $1.4 \times 10^{-18}$ for each system, as shown in Table 1. Both systems exploit in-vacuum, room-temperature thermal shields surrounding the lattice-trapped atoms (pictured in Fig. 1)[23]. This facilitates characterisation of a key systematic effect afflicting optical clocks, Stark effects from blackbody radiation bathing the ultracold atoms, and further provides an in-situ Faraday enclosure shielding the atoms from static electric fields due to stray charges on the vacuum apparatus[24]. These clocks also utilize a one-dimensional optical lattice operating near the magic wavelength, where lowest-order trap light shifts on the clock transition are cancelled, and higher-order lattice effects have been experimentally characterised[25]. Although alternative architectures utilize cryogenic operation[26] or three-dimensional optical lattices[27,28], the unprecedented clock performance reported here requires only a one-dimensional lattice and room-temperature operation, useful traits for future portable apparatus or robust primary standards of time and frequency. While we treat most systematic effects in the Methods, here we highlight two important effects that have not been experimentally characterised previously in optical clocks.

*Background gas collision shift.* Collisions with background gases lead to a shift of the atomic transition frequency. This shift is expected to scale with the trap loss rate, $\Gamma$, though this scaling has not been experimentally confirmed previously. Furthermore, theoretical determination of the scaling coefficient requires precise knowledge of the van der Waals coefficients of the two molecules[29]. The residual gas pressure in our two vacuum systems is ≈0.1 µPa, with $H_2$ by far the dominant gas species, determined by residual gas analysis. To measure the effects of background gas, we heat a non-evaporable getter pump and induce outgassing, decreasing the trap lifetime of Yb-1 to as low as 93 ms. Residual gas analysis confirms that the released gas is >95 % $H_2$. We determine the trap lifetime to better than 5 % uncertainty from the ratio of the atomic populations measured interleaving between two clock cycles with variable time delay before detection. We determine the induced shift by comparing the frequency of Yb-1 with Yb-2 serving as a stable frequency reference. As shown in Fig. 2a, the fractional frequency shift from background gas collisions is confirmed to vary linearly with loss rate, with a coefficient of $-1.64(12) \times 10^{-17}$ s. Characteristic values of the time constant for Yb-1 and Yb-2 are 3.0 s and 4.5 s, respectively, and this value is measured every few days. The shift for Yb-1 is therefore $-5.5(5) \times 10^{-18}$, and for Yb-2, $-3.6(3) \times 10^{-18}$. Uncertainty in the coefficient of the background gas shift is common-mode between the systems, and the differential uncertainty amounts to $3 \times 10^{-19}$, including both coefficient uncertainty and measurement uncertainty of the loss rate. We find that accurate determination of this shift is crucial for enabling $10^{-18}$ uncertainty with typical vacuum levels found in optical lattice clock systems. Additionally, repeating this measurement at several lattice depths, up to eight times its operational value (see Methods), we observe that this shift is independent of trap depth at the 10 % level.

*Spin-polarization impurity shift.* Collisions between lattice-trapped atoms give rise to a frequency shift. Due to the anti-symmetrisation condition of identical fermions, embodied in the Pauli exclusion principle, spin-polarized $^{171}$Yb atoms ($F = 1/2$ for the clock states) cannot interact with each other through even-partial-wave interactions (though we note that indistinguishability may be compromised by inhomogeneous probe excitation, see Methods). We achieve spin-polarization by optically pumping on the 556 nm transition in the presence of a 0.45 mT magnetic field[30]. If spin-polarization is incomplete, residual population remains in the depleted spin state, leading to imperfect suppression of *s*-wave collisional shifts. To quantify this effect, we intentionally compromise spin-polarization purity[31] by detuning the pumping laser frequency and measure frequency changes between high and low density. As shown in Fig. 2b, we find that the fractional frequency shift for the operational atom number, $N_0$, scales with $\eta$, the proportion of population in the depleted spin state, as $(5.50(12) \times 10^{-17}) \times \eta$.

The effectiveness of optical pumping into the $m_F = \pm 1/2$ state is evaluated by examining the maximal excitation ratio of the $\pi$-transition for atoms remaining in the depleted spin state, as displayed on Fig. 2c. For typical signal-to-noise ratios observed in our experiment, it is straightforward to verify that $\geq 99.5$ % of the atoms are prepared in the desired spin state. With this bound we constrain the *s*-wave collisional shift from imperfect spin-polarization to be $<3 \times 10^{-19}$ for Yb-1. The shift is $<1 \times 10^{-19}$ for Yb-2 due to the larger lattice beam waist (and correspondingly lower atom number density) afforded by the enhancement cavity (see Fig. 1). We note that the spin-1/2 nuclear structure, the simplest of any fermion, provides a consummate advantage to $^{171}$Yb in constraining the magnitude of this shift in comparison to other atomic species of interest. Failure to account for imperfect spin polarization can easily lead to errors above the $10^{-18}$ level.

**Instability.** Benefitting from the careful control of systematic shifts that compromise instability at long timescales, we demonstrate for the first time a clock instability that reaches into the $10^{-19}$ decade, as shown in Fig. 3. This level of performance is demonstrated in two complementary configurations: the blue dataset uses 560-ms Rabi spectroscopy, synchronized between Yb-1 and Yb-2, and the green dataset uses unsynchronized, 510-ms free-evolution-time Ramsey spectroscopy (see Methods). Taken over the course of 72 hours, with an uptime of 88 %, a single-clock measurement instability of $4.5 \times 10^{-19}$ is achieved in the former configuration, as determined by the final point of the total Allan deviation at 36 hours. An estimated measurement instability of $3.2 \times 10^{-19}$ is found by fitting a white frequency noise model to the total Allan deviation and extrapolating the fit to the full measurement time. Every 24 hours, the clocks were unlocked from the atomic transition so that an evaluation could be performed to ensure full compliance with the uncertainty budget of Table 1. During the clock comparison, the frequency difference is corrected in real-time for the blackbody shift (see Methods). A noise floor at $1 \times 10^{-18}$ is present for the uncorrected dataset, but the corrected data is consistent with white frequency noise throughout the entire set. This measurement stability demonstrates for the first time the possibility of geopotential determination approaching millimetre-level statistical resolution. We further note that this long-time performance does not require synchronized probing schemes, as a comparable instability of $6.4 \times 10^{-19}$ is obtained with the unsynchronized sequences (Fig. 3).

**Reproducibility.** Making use of the low measurement instability demonstrated, we undertake a campaign of frequency comparisons between the clocks to characterise the reproducibility of the two systems, as shown in Fig. 4. The frequency comparisons are performed with a blinding protocol (see Methods) that prevents the operator from having knowledge of the frequency difference during individual measurements. The results reported here are the culmination of ten blinded measurements, taken over the course of more than a month. A weighted average of these measurements leads to a frequency difference $(v_2-v_1)/v_{clock} = [-7\pm(5)_{stat}\pm(8)_{sys}] \times 10^{-19}$ after correcting for all relevant systematic effects. After four early comparisons, it was discovered that a faulty wire had removed the grounding connection to the conductive windows of Yb-2, compromising Faraday shielding and leading to a mid-$10^{-18}$ DC Stark shift between the systems. This experience underscores the indispensability of experimentally investigating reproducibility for substantiating an uncertainty budget.

**Discussion.**

We note that the gravitational redshift listed in Table 1 is similar for each clock, since the two systems are closely located, leading to a small relative uncertainty. However, the redshift transformation to the reference frame of TT has an uncertainty of $6 \times 10^{-18}$ (limited by the several-centimetre-level state-of-the-art geodetic determination), much larger than the total measurement uncertainty between the clocks[32]. In other words, if these clocks were compared across a long baseline or used for remote comparisons with other clocks around the world, the measurement would be limited by gravitational knowledge on Earth's surface. Refinements of geoid models using satellite-based long-wavelength data[33] and terrestrial short-wavelength data[34] may reduce future geoid model uncertainty to a degree, but height uncertainty at or below 1 cm (fractional frequency uncertainties below $1 \times 10^{-18}$) remains at best an optimistic goal[21,32]. With optical clock performance at the levels demonstrated here, these clocks now enable beyond-state-of-the-art geodetic measurements and fundamental physics studies[7,8,10,13,15].

## Methods.

**Experimental setup.** Ytterbium-171 from an effusive cell is slowed and cooled to millikelvin temperature by a three-dimensional magneto-optical trap operating on the allowed $^1S_0 \rightarrow ^1P_1$ transition at 399 nm. The atoms are further cooled to a few microkelvin by a three-stage 556 nm magneto-optical trap on the spin-forbidden $^1S_0 \rightarrow ^3P_1$ transition. The ultracold atomic sample is loaded into an optical lattice close to the magic wavelength, $\lambda_{latt} = 759$ nm. The lattice of Yb-1, with a $1/e^2$ power radius of 75 μm, is formed by retroreflecting the laser upon itself. A larger radius of 170 μm is achieved in Yb-2 by means of an enhancement cavity[25]. Both lattices operate at a trap depth of 50 $E_r$. The lattice photon recoil energy is given by $E_r = h^2/(2m\lambda_{latt}^2)$, where $h$ is Planck's constant and $m$ is the mass of $^{171}$Yb. The atoms are then cooled to a longitudinal temperature of 500 nK by 20 ms of sideband cooling on the $^1S_0 \rightarrow ^3P_0$ transition at 578 nm, quenched by the $^3P_0 \rightarrow ^3D_1$ transition at 1,388 nm[35,36]. The atoms are spin-polarized to >99.5 % purity on the 556 nm transition. We elect to operate with an atom number, $N_0$, of 5,000 lattice-trapped atoms for both clocks.

Rabi spectroscopy with an interrogation time of 560 ms is performed on the ultra-narrow (natural linewidth of 7 mHz), doubly-forbidden 578 nm line, leading to a Fourier-limited full width at half maximum of 1.4 Hz, shown in Fig. 1. By applying a bias field of 0.1 mT, the Zeeman spectral lines are split by 400 Hz, and locking is performed as described in ref. 30. The clock laser is stabilized to a 29-cm cavity made of ultra-low expansion glass with an instability assessed as $\leq 1.5 \times 10^{-16}$ by using the atoms as a frequency discriminator[4]. After spectroscopy, the normalized excitation ratio is detected by collecting fluorescence on the $^1S_0 \rightarrow ^1P_1$ transition and repumping on the $^3P_0 \rightarrow ^3D_1$ transition. The total cycle time is 860 ms, leading to a probe time to cycle time ratio of 65 %. An atomic shutter blocks the atomic beam during spectroscopy, preventing collisional shifts from the atomic beam as well as excess blackbody radiation shifts from the hot oven. The measurements reported here are taken in two complementary configurations: 1) frequency corrections are sent independently to AOM-1 and AOM-2 in Fig. 1, seen by Yb-1 and Yb-2 respectively; and 2) corrections from Yb-1 are sent to AOM-0, seen identically by both systems. In case 1, the frequency difference is simply given by the difference between the two lock integrators. In case 2, the frequency difference is inferred by using Yb-2 as a frequency discriminator. Four measurements, averaging to $-5(6) \times 10^{-19}$, are taken in configuration 1. Six measurements, averaging to $-8(7) \times 10^{-19}$, are taken in configuration 2.

**Cold collisional shifts.** While $s$-wave interactions between identical $^{171}$Yb atoms are highly suppressed by the Pauli exclusion principle, $p$-wave collisions are allowed[37]. However, with $l \geq 1$ angular momentum, a potential barrier forms from the competition of van der Waals attraction and centrifugal repulsion[38]. As a result, such collisions start to freeze out below the $p$-wave barrier, calculated to be 30 μK[39]. Our trap depth of 50 $E_r$ corresponds to a maximum atomic temperature of 4.8 μK, and we thus expect collisional shifts to be largely frozen out. Inhomogeneous probe excitation may lead to a residual $s$-wave collisional shift[40], though previous measurements indicate that this effect is usually small compared to $p$-wave interaction shifts, even at typical ultracold temperatures[37]. To aid in the measurement of collisional shifts at low trap depths, atom number is enhanced by quenched sideband cooling of atoms trapped in a deeper lattice (200 $E_r$), followed by an adiabatic ramp to 50 $E_r$ depth. Motional sideband spectroscopy is performed, and no significant difference is observed in the motional state distribution between this population and a sideband-cooled population loaded directly into a 50 $E_r$ lattice. By evaluating the density-dependent collision shift at a range of atom numbers and trap depths, the cold collision shift is measured to be only $-2.1(7) \times 10^{-19}$ for the operational atom number, $N_0$, of Yb-1. For Yb-2, the cold collision shift is smaller still, -0.4(2) $\times 10^{-19}$, because of the larger beam waist present in the enhancement cavity. Given the small magnitude of the observed density shift, we note that even operating at an atom number of 200,000, corresponding to a one-second quantum-projection-noise of $<2 \times 10^{-18}$, leads to a Yb-2 collisional shift uncertainty in the $10^{-19}$ decade.

**Doppler shift.** By operating in the Lamb-Dicke regime, the atoms of an optical lattice clock are essentially stationary with respect to the lattice[41], but motion of the lattice with respect to the laboratory frame results in a first-order Doppler shift. This shift can be explored by introducing a delay before spectroscopy, thus altering the phase and amplitude of the motion experienced during spectroscopy. Uncontrolled, we have observed first-order Doppler shifts in excess of $1 \times 10^{-16}$ for previous generations of our clock. Lattice motion with respect to the clock laser frame can be reduced by employing clock laser phase-noise cancellation with the lattice retroreflector as a phase reference[42,43], though it becomes important to understand how completely this technique can afford suppression. Without cancellation, we evaluate this shift to be $\leq 1 \times 10^{-16}$ for Yb-1 and $\leq 0.5 \times 10^{-16}$ for Yb-2 under the present operating conditions.

To evaluate the fidelity by which the clock laser is coherently transferred to the lattice reference frame, we intentionally induce a large first-order Doppler shift. This is done by sweeping the voltage of an electro-optic modulator in the clock laser path, leading to an optical path length variation rate ranging from 30 μm/s to 1,500 μm/s. After engaging phase-noise cancellation to the lattice reference frame, we observe that this shift is suppressed by a factor of $\geq 5,000$. Suppression is complete for all optical path length variation rates that are investigated. Applying this suppression factor to the uncancelled Doppler shift, we assess this shift to be no larger than $2 \times 10^{-20}$ for Yb-1 and $1 \times 10^{-20}$ for Yb-2.

**BBR Stark shift.** By far the largest uncancelled shift present in our [171]Yb clock is an AC Stark shift due to blackbody radiation (BBR) from the clock's room-temperature environment. This shift is characterised at better than the part-per-thousand level by means of an in-vacuum thermal shield that provides the atoms with a near-ideal blackbody environment. This apparatus has been evaluated comprehensively and is summarized by Table I of ref. 23, with the following improvements. The resistive thermal detectors are now mounted in a manner that better preserves manufacturer calibration, eliminating post-calibration fidelity as a source of error. The detectors are packed with more aluminium oxide fillings that facilitate heat transfer with the shield in vacuum, reducing self-heating. We have also evaluated inhomogeneous window heating due to lattice laser absorption, finding that for Yb-2 the enhancement cavity increases the effective temperature of the windows on the vertical axis by 440 mK, leading to a shift correction of $6(3) \times 10^{-19}$. For Yb-1, lattice heating is not significant at the mid-$10^{-19}$ level. With these modifications, the total environmental shift uncertainty is $4 \times 10^{-19}$ and $5 \times 10^{-19}$ for Yb-1 and Yb-2, respectively.

The temperature of the shield is determined by five platinum resistive thermal detectors: three mounted on the body of the shield, one on the top window and one on a side window. During the frequency comparisons, we measure each of the thermal detectors every 100 s and apply a correction to each clock frequency based upon the effective temperature of the system, with effective temperature defined as in ref. 23. We note that the three thermal detectors on the body agree to within the calibration uncertainty of <5 mK. Even in the tightly controlled thermal environment of the laboratory, real-time corrections are necessary to remove frequency drift between the clocks. In Fig. 3 it is seen that without correcting for the BBR shift, the clocks encounter a noise floor of $1 \times 10^{-18}$, consistent with the ~100 mK deviations we observe. With BBR corrections properly applied, no noise floor is observed and the clocks average below $5 \times 10^{-19}$.

The greater part ($8.5 \times 10^{-19}$) of the BBR shift uncertainty is due to uncertainty in the coefficient of the so-called dynamic correction to atomic response[44]. Improved measurement of the $^3D_1$ lifetime and branching ratio to $^3P_0$ can reduce this source of uncertainty. Atomic response is common-mode between the two room-temperature clocks, but uncertainty due to blackbody environment is uncorrelated. The differential uncertainty is therefore a quadrature sum of the environmental uncertainty, amounting to $6 \times 10^{-19}$.

**Lattice light shift.** By operating at the magic wavelength, where the electric dipole (E1) polarizabilities of the ground and excited states precisely cancel, clock frequency can be made largely insensitive to lattice laser intensity[41]. This technique allows much better than part-per-million cancellation of the lattice AC Stark shift, but at the current frontier of optical clock performance higher order polarizabilities due to e.g., M1-, E2- and two-photon E1-transitions are relevant. Combined with motional state quantization in the lattice, these effects introduce a non-polynomial dependence on intensity, as well as dependence on atomic temperature[45]. Recent analysis has found that the scaling of atomic temperature with trap depth can lead to a simplification of the shift, allowing it to be modelled as a polynomial series[25]. A further implication of these effects is that the E1-polarizability affects higher-order trap-depth-dependent terms, leading to meaningful frequency-dependence of these terms[46]. For the 50 $E_r$ lattice trap depth we employ here, we find that a cubic fit is sufficient to model relevant light shifts with an error $\leq 3 \times 10^{-19}$. We note that the model uncertainty of the lattice light shift is common-mode between the two clocks and is thus suppressed in the differential uncertainty.

If the forward-going and backward-going lattice beams of Yb-1 do not have the same intensity, a travelling wave is present, leading to an apparent shift of the power series coefficients[26]. Allowing for an intensity mismatch of up to 15 % due to scattering, absorption or imperfect retroreflection and focusing, we conservatively constrain the travelling wave contamination of Yb-1 to be no more than $1 \times 10^{-19}$. The enhancement cavity prevents any significant lattice intensity mismatch from occurring on Yb-2.

To minimize the negative effects of the lattice light shift, we choose to tune our lattice laser to the operational magic wavelength, at which a positive quadratic shift partially cancels a negative linear shift, yielding a shift insensitive to changes in trap depth in the vicinity of our operational trap depth[45]. For our sideband-cooled sample, we find an operational magic wavelength of 394,798,267.7(5) MHz, leading to an experimental lattice light shift uncertainty of $8 \times 10^{-19}$. The differential uncertainty is only $2 \times 10^{-19}$, as the operational magic wavelength suppresses uncertainty due to small differences in trap depth. The lattice laser is stabilized to a cavity made of ultra-low expansion glass and is measured on a weekly basis with a resolution better than 10 kHz by an octave-spanning Ti:sapphire optical frequency comb referenced to a calibrated hydrogen maser.

**Zeeman shift.** As the clock is referenced to a transition between spherically symmetric ($J = 0$) electronic states, the atomic wave function is insensitive at first order to any non-scalar effect, such as coupling to a magnetic field. Due to non-zero nuclear spin, a small degree of linear field sensitivity is present. The difference in splitting between $\pi$- and $\sigma$-transitions can be used to measure the magnetic field, and thus the linear Zeeman sensitivity can be determined[47]. We find that the linear Zeeman effect splits the $\pi$-transitions by 199.516(2) Hz/G from centre, in good agreement with a previous measurement[30].

By iteratively interrogating the two nuclear spin states and taking the average, the first-order Zeeman shift is cancelled completely. We confirm this experimentally to the $4 \times 10^{-19}$ level by measuring the clock frequency difference between large magnetic fields with opposite polarity. The measured splitting yields a readout of the magnetic field, used to precisely determine the second-order Zeeman coefficient. By interleaving between two clocks at differing magnetic fields, the coefficient is found to be -0.06095(7) Hz/G$^2$, in good agreement with a previous measurement[37]. Precise spectroscopic determination of the magnetic field is complicated by the existence of a polarization-dependent vector Stark shift from the lattice laser that acts as a pseudo-magnetic field[47]. In practice, it is straightforward to reduce this shift to <100 mHz by utilizing a linearly polarized lattice laser. For Yb-1, the second-order Zeeman shift is -118.1(2) $\times 10^{-18}$, with an uncertainty limited by knowledge of the residual vector Stark shift. For Yb-2, a well-defined linear eigen-polarization is coupled into the build-up cavity, leading to a negligible vector Stark shift[25] and a Zeeman uncertainty of only $1 \times 10^{-19}$, limited by knowledge of the second-order coefficient. The vector Stark shift is uncorrelated, but the coefficient uncertainty is common-mode, yielding a differential uncertainty of $1 \times 10^{-19}$.

**DC Stark shift.** Stray DC electric fields have been observed to cause Stark shifts on the order of $10^{-13}$ for some optical lattice clocks[48]. Our thermal shield serves as a Faraday cage to null stray electric fields. During normal clock operation, the body of the shield and the windows, coated with conductive indium-tin oxide, are grounded. By applying voltage to the windows of the thermal shield, the Stark shift due to stray electric fields is found to be consistent with zero at better than the $10^{-19}$ level on both systems[24].

**Probe Stark shift.** A measurement of the probe AC Stark shift, arising from the clock light itself, is performed by measuring the clock shift between normal operation and a case where the clock laser is phase-modulated, and the clock transition is excited by a weak resonant sideband 40 dB lower intensity than the off-resonant carrier. For 560-ms Rabi spectroscopy, the probe shift is found to be $2 \times 10^{-20}$, with uncertainty conservatively assessed at 50 % of its value. This shift is largely removed in common-mode for the differential measurement.

**Line pulling.** The presence of other spectroscopic features near the atomic transition can lead to an apparent shift of the line. $\sigma_\pm$-transitions result from driving the $\Delta m_F = \pm 1$ transition. These transitions, detuned from the clock transition by 1.15 kHz, are typically suppressed to about 1 % by using linearly polarized clock light parallel to the quantization axis of the atoms. Imperfect spin polarization might lead to a residual population of clock atoms in the depleted nuclear spin state. This population is suppressed by at least 99.5 %, and the $\pi$- ($\sigma$-) transitions are detuned by $\pm 400$ ($\mp 750$) Hz. The closest spectroscopic features are the transverse sidebands, detuned by only 60 Hz from the carrier. However, they are highly symmetric due to the large number of transverse motional states and are largely suppressed by probing collinear to the lattice laser. Due to a relatively narrow Fourier-limited linewidth of 1.4 Hz (Fig. 1), we place a conservative upper bound of $1 \times 10^{-19}$ on all sources of uncertainty from line pulling.

**Tunnelling.** In a horizontal lattice, tunnelling of atoms between lattice sites could potentially result in a shift as large as the bandwidth of the Bloch state. By orienting the lattice vertically, adjacent lattice sites are energetically nondegenerate by $h \times \Delta_g = mg(\lambda_{latt}/2)\cos\theta \approx h \times 1.6$ kHz, where $g = 9.8$ m/s$^2$ is the acceleration due to gravity and $\theta \approx 1°$ is the lattice's declination from vertical. This nondegeneracy induces atomic localization through periodic Bloch oscillations, spectroscopically manifest as sidebands corresponding to transitions between Wannier-Stark states. It has been shown theoretically that interference between the carrier and Bloch sidebands can lead to a shift of order of magnitude $\Omega_0\Omega_1/\Delta_g$, where $\Omega_0$ and $\Omega_1$ are the Rabi frequencies of the carrier and first-order Bloch sideband[49]. By investigating the amplitude of the Bloch sideband for a typical sample at 500 nK (corresponding to a mean longitudinal motional number $<n_z> = 0.04$), it is found that $\Omega_1 = (3 \times 10^{-4})\Omega_0$. For 560-ms Rabi spectroscopy, we conservatively assess the tunnelling shift as $<1 \times 10^{-21}$.

**Servo error.** Local oscillator frequency drift can result in a locked frequency offset from the atomic transition, due in part to the delay between atomic interrogation and frequency feedback correction. We control servo error by probing our transition first in ascending frequency order followed by descending frequency order, cancelling cavity drift at <500 μHz/s by digital feed-forward correction and implementing a second integrator. By analysis of the lock error signal during extended clock operation, we observe servo error consistent with zero at the sub-$10^{-19}$ level for our longest runs. Due to the shared local oscillator, servo error is common-mode between the two clocks, and is suppressed to a high degree.

**Optical frequency synthesis shifts.** Clock operation requires pulsing the 578 nm interrogation laser with an acousto-optic modulator, a process which is known to induce phase chirps. The switching AOM also serves as the phase-noise-cancellation AOM[42] that, when locked, leads to phase transients that are largely random, averaging to <50 mrad, and are suppressed with a time constant of <20 μs. Following the analysis of ref. 43, we place a conservative upper bound of $1 \times 10^{-20}$ on this source of uncertainty.

Updating the frequency of a direct digital synthesizer (DDS) can also lead to phase transients that are seen as a frequency shift on the clock transition[50]. By synchronizing a counter to the clock interrogation and directly counting the DDS frequency, we confirm that this technical source of error is consistent with zero at the level of $7 \times 10^{-20}$, leading to a differential uncertainty of $1 \times 10^{-19}$

**Determination of geopotential.** As dictated by relativity, measurements of time in different reference frames should be transformed into a shared reference system. A clock that is elevated above the geoid experiences a gravitational blueshift, as well as a redshift due to increasing centrifugal acceleration[51]. A recent state-of-the-art determination of the geopotential of the Q407 marker on the NIST-Boulder campus found a clock shift of $(179,853(6)) \times 10^{-18}$ from the reference frame of TT[32]. Spirit levelling paired with high-accuracy gravimetry is used to determine a further shift of $(810.9(2)) \times 10^{-18}$ between the Q407 marker and the laboratory floor. The atomic sample is localised by means of the position of the atomic detection laser, and the heights of the lattice-trapped atoms within Yb-1 and Yb-2 are measured to 2 mm resolution, leading to a further shift of $154.9(2) \times 10^{-18}$ and $151.1(2) \times 10^{-18}$, respectively.

**Synchronized operation.** The main focus of our study is the characterisation of systematic effects and of the level at which they can be controlled. We therefore choose to perform clock comparisons with synchronous interrogation to enhance measurement instability by rejecting the Dick effect in common mode[52,53]. Despite rejecting the Dick effect, synchronous interrogation does not reject other sources of error, and reaching low instability at long timescales requires control of all systematic shifts. To demonstrate the frequency stability afforded by the clocks, we measure instabilities at the $10^{-19}$ level for both synchronized and unsynchronized modes of operation, as shown in Fig. 3. For the unsynchronized measurements, we utilize Ramsey spectroscopy with a free-evolution-time of 510 ms so that the enhanced quality factor enables a one-second instability similar to the synchronized Rabi case. We note that the one-second instability of $1.5 \times 10^{-16}$ remains significantly higher than the expected quantum-projection-noise limit ($< 5 \times 10^{-17}$, for the conditions discussed above), due to technical sources of noise.

**Blinding protocol and selection criteria.** Operator bias can be an important source of error in any physical measurement. For example, for frequency comparisons such as those reported in Fig. 4 it would be possible for the operator to stop the experiment when statistical fluctuations temporarily push the average frequency difference to zero. To eliminate this source of bias, we employ a blinding protocol that adds a large offset to the frequency difference observed by the operator. This offset is pseudorandomly chosen from a uniform distribution spanning ±1 kHz, many orders of magnitude greater than the sub-millihertz resolution of the experiment. The operator is completely blind to the frequency difference until the termination of the measurement.

Prior to commencing each blinded measurement, both clocks are comprehensively assessed to verify that they are fully compliant with the uncertainty budget of Table 1. This assessment consists of a checklist of all parameters that can change from day to day. For instance, the operator verifies via motional sideband spectroscopy that the trap depth is within 10 % of 50 $E_r$ and that the longitudinal temperature is lower than 1 μK. A post-selection criterion, that data averages as white frequency noise for averaging times greater than 100 s, is also established. An eleventh dataset, with a noise floor of $7 \times 10^{-18}$ at $\geq 10^4$ s, is excluded because of the latter criterion. We note that, due to the larger error bars associated with it, inclusion of this set would not significantly change the average. With no omissions, each of the ten blinded measurements subject to these criteria is included in the computation of the average frequency difference of -7(5)$_{stat} \times 10^{-19}$. Statistical error bars are assigned by applying a $1/\sqrt{\tau}$ fit to the total Allan deviation assuming a servo attack time of 20 s and extrapolating to the full measurement time.

**Data availability.** The data that support the findings of this study are available from the corresponding author upon reasonable request.

**Acknowledgements.**

The authors acknowledge financial support from the National Institute of Standards and Technology, The NASA Fundamental Physics programme, the Defense Advanced Research Projects Agency (DARPA) Quantum Assisted Sensing and Readout (QuASAR) programme and PECASE. R.C.B. acknowledges support from the National Research Council Research Associateship programme. We also thank T. Fortier and H. Leopardi for femtosecond optical frequency comb measurements, and J. Kitching and D. Hume for careful reading of this manuscript.

**Author contributions.**

W.F.M., X.Z., R.J.F., S.A.S., D.N. and A.D.L. carried out the instability and reproducibility measurements. W.F.M., X.Z., S.A.S., K.B., D.N., R.C.B., N.H., G.M., M.S., T.H.Y. and A.D.L. contributed to the evaluation of the uncertainty budget. A.D.L. supervised this work. All authors contributed to the final manuscript.

**Additional information**

Reprints and permissions information is available online at www.nature.com/reprints. Correspondence should be addressed to A.D.L. (email: andrew.ludlow@nist.gov). Contributions to this article by workers at NIST, an agency of the U.S. Government, are not subject to copyright.

**Competing financial interests.**

The authors declare no competing financial interests.

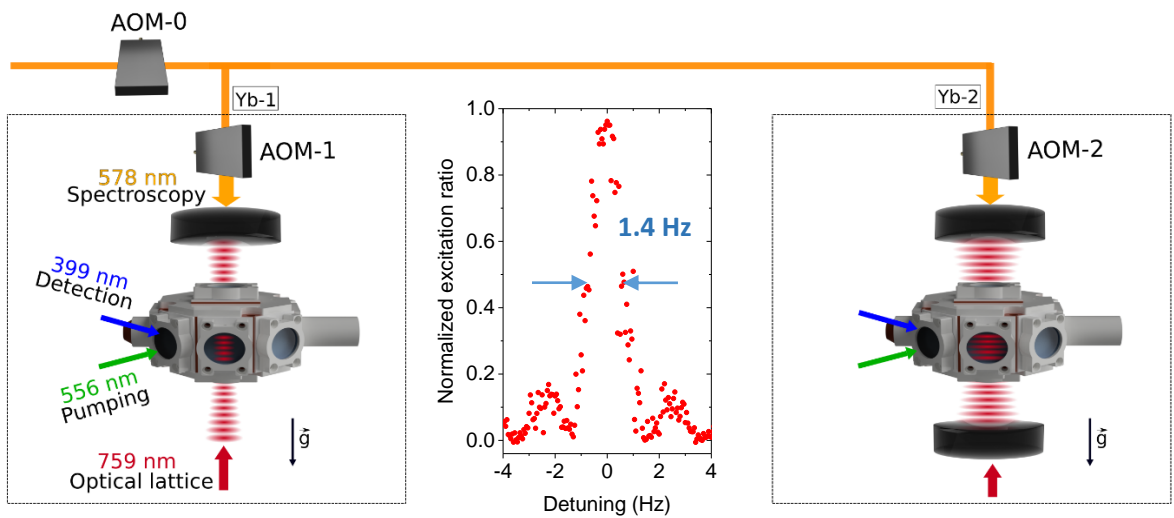

**Figure 1 | Simplified experimental scheme.** [171]Yb atoms are cooled and loaded into two vertically oriented one-dimensional optical lattices. Frequency corrections are applied by three acousto-optic modulators (AOMs). AOM-1 and AOM-2 also cancel the optical path length fluctuations for Yb-1 and Yb-2, with the lattice mirror serving as a phase reference. The pumping laser creates a spin-polarized atomic sample, and the detection laser reads out atomic populations. The atoms are surrounded by an in-vacuum room-temperature thermal shield (see Methods). The inset displays the typical 560 ms Rabi spectrum with a Fourier-limited full width at half maximum of 1.4 Hz. Shown is a single trace with no averaging and signal-to-noise ratio characteristic of the measurements reported here.

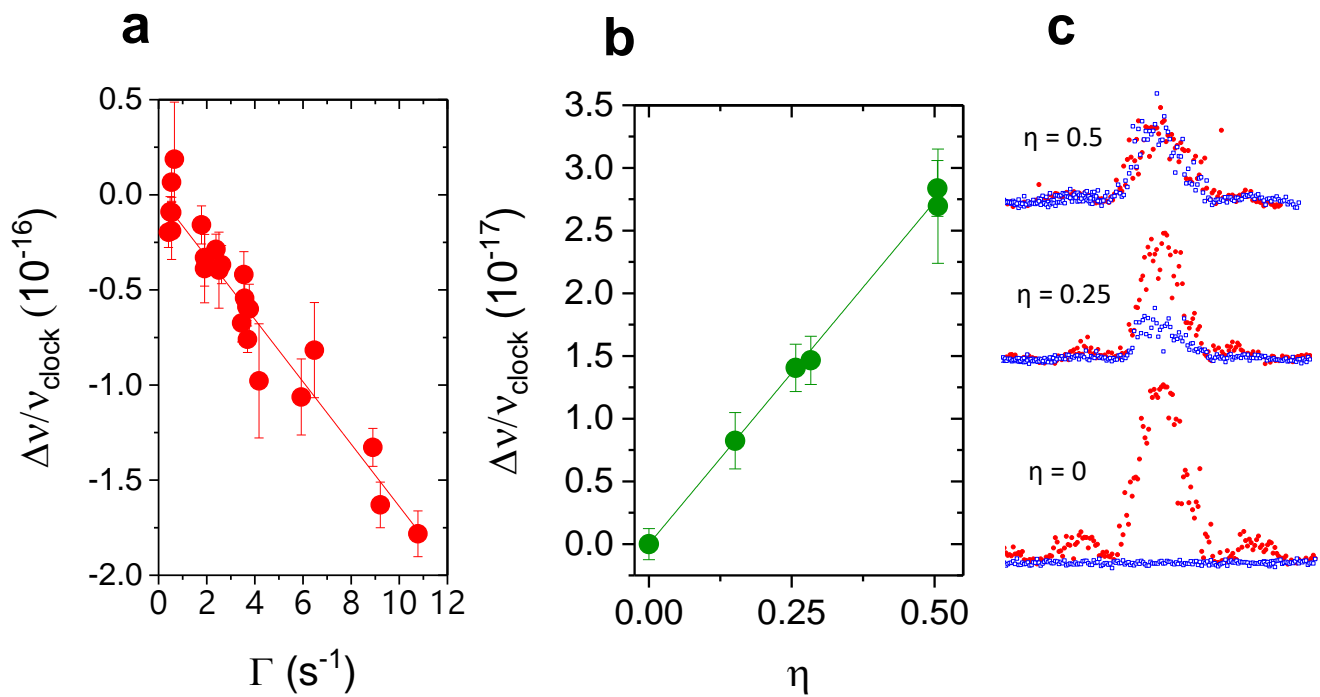

**Figure 2 | Sources of systematic uncertainty. (a)** Measurement of shifts due to collisions with background gas molecules. A fit to the dataset yields a shift of $\Delta\nu/(\nu_{clock}\,\Gamma) = (-1.64(12) \times 10^{-17})$ s. Error bars are the last point on the total Allan deviation for frequency comparisons (representing half the measurement time of approximately one hour) between Yb-1 and Yb-2. **(b)** Cold collisional shift measurements, as a function of spin polarization impurity. A fit to the dataset yields a shift of $\Delta\nu/(\nu_{clock}\,\eta) = 5.50(12) \times 10^{-17}$ for atom number, $N_0 \approx 5,000$ atoms. Error bars are the last point on the total Allan deviation after a measurement time of approximately one hour. **(c)** Traces over the Rabi line for variable degrees of optical pumping. Red filled circles are the $\pi$-transition corresponding to the desired spin state, and blue empty squares are the undesired spin state. From bottom to top, the insets correspond to a detuning of 0 kHz, 680 kHz, and 6,800 kHz from the optical pumping transition (natural linewidth = 180 kHz).

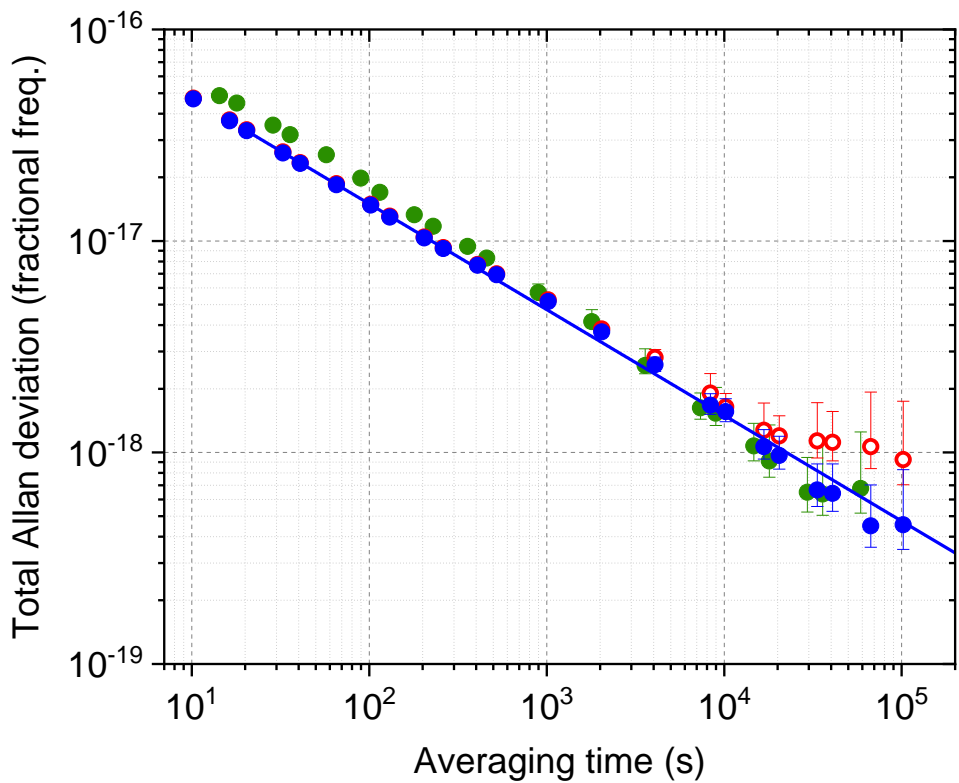

**Figure 3 | Measurement instability.** Total Allan deviation representing an upper bound on the single-clock measurement instability, $\left(\frac{1}{\sqrt{2}}\right)[\nu_2(t) - \nu_1(t)]/\nu_{clock}$, and its white frequency noise asymptote of $1.5 \times 10^{-16}/\sqrt{\tau}$, where $\tau$ is the averaging time in seconds. The blue and red sets use synchronized Rabi spectroscopy. The green set uses unsynchronized Ramsey spectroscopy. Blue and green circles are line-by-line corrected for the blackbody shift (see Methods); red circles are uncorrected. Error bars represent the 1σ uncertainty in the Allan deviation.

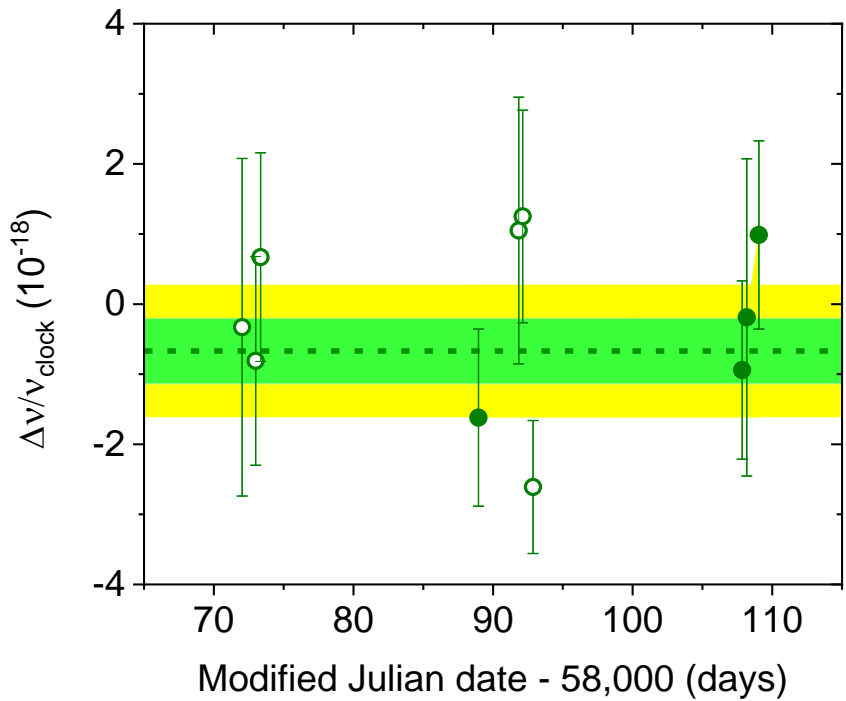

**Figure 4 | Reproducibility.** Green filled (empty) circles represent blinded measurements taken in configuration 1 (2), as described in the Experimental setup section of Methods. All systematics have been subtracted from the measurements. The dotted line represents the mean, and the green (yellow) shaded region represents the statistical (total) 1σ uncertainty of the measurements; $\Delta\nu/\nu_{\text{clock}} = [-7\pm(5)_{\text{stat}}\pm(8)_{\text{sys}}] \times 10^{-19}$. The statistical uncertainty is scaled up by the square-root of the reduced chi-squared statistic, $\chi^2_{red} = 1.06$. Error bars represent 1σ uncertainty obtained from the two-clock total Allan deviation, as described in Methods.

| Shift ($10^{-18} \times \nu_{\mathrm{clock}}$) | Yb-1 Shift | Yb-1 Uncertainty | Yb-2 Shift | Yb-2 Uncertainty | Differential Uncertainty | |
|---|---|---|---|---|---|---|
| Background gas collisions | -5.5 | 0.5 | -3.6 | 0.3 | 0.3 | |
| Spin polarization | 0 | <0.3 | 0 | <0.1 | <0.3 | |
| Cold collisions* | -0.21 | 0.07 | -0.04 | 0.02 | 0.07 | |
| Doppler | 0 | <0.02 | 0 | <0.01 | 0.02 | |
| Blackbody radiation* | -2,361.2 | 0.9 | -2,371.7 | 1.0 | 0.6 | |
| Lattice light (model) | 0 | 0.3 | 0 | 0.3 | <0.1 | |
| Travelling wave contamination | 0 | <0.1 | 0 | <0.01 | <0.1 | |
| Lattice light (experimental) | -1.5 | 0.8 | -1.5 | 0.8 | 0.2 | |
| Second-order Zeeman* | -118.1 | 0.2 | -117.9 | 0.1 | 0.1 | |
| DC Stark | 0 | <0.07 | 0 | <0.04 | <0.08 | |
| Probe Stark | 0.02 | 0.01 | 0.02 | 0.01 | <0.01 | |
| Line pulling | 0 | <0.1 | 0 | <0.1 | <0.1 | |
| Tunnelling | 0 | <0.001 | 0 | <0.001 | <0.001 | |
| Servo error | 0.03 | 0.05 | 0.03 | 0.05 | <0.01 | |
| Optical frequency synthesis | 0 | <0.1 | 0 | <0.1 | <0.1 | |
| **Total** | **-2,486.5** | **1.4** | **-2,494.7** | **1.4** | | |
| | | | | | | |
| Gravity shift from TT reference frame | 180,819 | 6 | 180,815 | 6 | 0.3 | |
| **Total shift from TT reference frame** | **178,333** | **6** | **178,320** | **6** | **0.8** | |

**Table 1 | Characteristic clock uncertainty budget.** All values are in units of $10^{-18} \times \nu_{\mathrm{clock}}$. In the differential measurements, some effects are removed in common-mode between the clocks, while others are uncorrelated, leading to a total uncertainty smaller than either individual clock. Description of each of these effects can be found in the main text and Methods.

*Shifts are calculated and corrected in real-time. Listed values represent the average shift for a typical run.